\documentclass[]{llncs}

\usepackage[]{textcomp}
\usepackage[]{microtype}
\usepackage[utf8]{inputenc}
\usepackage[T1]{fontenc}
\usepackage[]{stmaryrd}
\usepackage[]{latexsym}

\usepackage{hyperref}

\let\oldampersand\&

\renewcommand*\&{{\itshape\oldampersand}}

\title{The tree machine\thanks{This research has received funding from the European Research Council under the FP7 grant agreement 278673, Project MemCAD}}

\author{Arnaud Spiwack}

\institute{MINES ParisTech\\
\email{arnaud@spiwack.net}}

\begin{document}
  \maketitle
  \begin{abstract}
    A variant of Turing machines is introduced where the tape is replaced by a single tree which can be manipulated in a style akin to purely functional programming. This yields two benefits: first, the extra structure on the tape can be leveraged to write explicit constructions of machines much more easily than with Turing machines. Second, this new kind of machines models finely the asymptotic complexity of functional programming languages, and may allow to answer questions such as ``is this problem inherently slower in functional languages''.
  \end{abstract}
  \section{Intro}
  \par
  This article came to be as I seem to find myself all to often in two kinds of discussions: one of them is the functional programmer's complaint that Turing machine make an unpleasant computation model as it is so unstructured that writing any explicit Turing machine is a chore usually left to the gods of hand-waving. The second one is a common interrogation about some computational problem: ``is it actually slower by a logarithmic factor to solve with a purely functional program, rather than an imperative one''.
  \par
  My inner functional programmer's reflex would be to turn to ${\lambda}$-calculus to answer such questions. However, there is no denying that it is easier to quantify over automata-like machines, such as Turing-machines, for the purpose of proving complexity results. With that in mind, I will introduce, in this article, a variant of Turing machines, the tree machine, with better structured data which correspond faithfully to the cost-model of purely functional programming languages.
  \par
  I will not attempt to answer, either positively or negatively, whether concrete problems are slower or not in purely functional style; I will however, give an explicit description of a machine implementing ${\lambda}$-calculus in section~\ref{s:lambda}, to demonstrate that it is effectively possible to write non-trivial machines explicitly in this model.
  \par
  \paragraph{Acknowledgment} I want to thank Alexandre Miquel and Guyslain Naves who, most independently, planted the seed of this article in my mind through very entertaining and enlightening discussions.\section{Eilenberg's machines}
  \par
  We will work with a generic notion of machines introduced by Eilenberg~\cite[Chapter 10]{Eilenberg1974}, which can be instantiated to yield finite automata as well as Turing machines. The tree machine introduced in Section~\ref{s:treemachine} is yet another instantiation of Eilenberg's machine (in fact we will give several equivalent definitions).
  \par
  A type of machine is given by a set $X$ of \emph{data} and a set $\Phi \subseteq \mathcal{P}{\left( X\times X\right) }$ of \emph{instructions}. Most of the times the instructions will be partial relations. A machine of type $\left( X,\Phi \right) $ is given by a finite set $Q$ of \emph{states}, and subsets $I$ and $F$ of initial and finite states, as usual for automata. Transitions are labelled with relations of $\Phi $. A path ${q}_{0}$,{\ldots},${q}_{n}$ computes the composition of the relations on the successive edges. The machine itself compute the union of the relation computed by path from an initial state to a final state.
  \par
  It does not change the expressiveness to close $\Phi $ by composition (${i}_{1}\cdot {i}_{2}$), union (${i}_{1}+{i}_{2}$), identity ($\mathbf{1}$)\footnote{In Eilenberg's formulation, a more general kind of relation is considered, in order to be able, typically, to count the multiplicity of successful path. In that case, closure by $\mathbf{1}$ (\emph{i.e.} adding ${\epsilon}$-transitions) is not permitted.} and empty relation ($\mathbf{0}$). We shall use this fact implicitly.
  \par
  In fact, relations computed by a machine of type $\left( X,\Phi \right) $ are exactly the relations in the sub-Kleene algebra of $\mathcal{P}{\left( X\times X\right) }$ generated by $\Phi $: the result for finite automata lifts naturally to Eilenberg's machines. So Eilenberg machines are equivalent to regular expressions with alphabet $\Phi $. However, if closing $\Phi $ by all the regular expression operations does not change what relation the machines compute, it does change the complexity. Since we are concerned with complexity properties, we may therefore refer to regular expressions as machines, while we will reserve the term \emph{instructions} to star-free expressions.
  \par
  This definition of Eilenberg's machines is naturally non-deterministic. It would be more accurate to work with deterministic machines in the setting of this article, but it does not really change anything of substance, and would unnecessarily clutter the presentation. So the machines throughout this article will be non-deterministic, but all of them could be made deterministic, and actually should, for practical applications.
  \section{Tree machines}\label{s:treemachine}
  \par
  Let us define the set $T$ of (rooted, unlabeled, binary) trees as the set generated by the following grammar:
  \begin{displaymath}
    \begin{array}{l@{~::=~}l@{~\mid ~}l}
      \mbox{$u$,$v$} & \left( \right)  & \left( u,v\right) \\
    \end{array}
  \end{displaymath}
  Such trees will be the data of our tree machines. Take notice of the fact that trees do not replace the alphabet of Turing machines but the whole tape: there will not be a tape of trees, just one tree.
  \par
  Let us define the following partial functions on $T$:
  \begin{displaymath}
    \parbox{0.85\linewidth}{$
    \begin{array}{l@{\quad }c@{\quad }l@{\qquad }l}
      \delta {\left( x\right) } & = & \left( x,x\right)  & \\
      {\pi }_{1}{\left( x\right) } & = & y & (\mbox{if $\exists {z}^{\in T}.\,\,x=\left( y,z\right) $})\\
      {\pi }_{2}{\left( x\right) } & = & z & (\mbox{if $\exists {y}^{\in T}.\,\,x=\left( y,z\right) $})\\
      \left( i,j\right) {\left( x\right) } & = & \left( i{\left( y\right) },j{\left( z\right) }\right)  & (\mbox{if $x=\left( y,z\right) $, for $i$ and $j$ partial functions})\\
      \varepsilon {\left( x\right) } & = & \left( \right)  & \\
      \left( \right) {\left( x\right) } & = & \left( \right)  & (\mbox{if $x=\left( \right) $})\\
    \end{array}
    $}
  \end{displaymath}
  \par
  The set of instruction $\Phi $ is chosen to be the smallest set containing the partial functions $\left\{ \delta ;{\pi }_{1};{\pi }_{2};\varepsilon ;\left( \right) ;\mathbf{1}\right\} $ and closed by $\left( \cdot ,\cdot \right) $. This set of instructions has been chosen to correspond to the presentation of cartesian products and terminal elements in categories as adjunctions.
  \par
  We call \emph{tree machine} a machine of type $\left( T,\Phi \right) $. Notice that, contrary to Turing machines, tree machines are not parametrised by an alphabet: the tree structure offers enough power on its own.
  \par
  Tree machines, by virtue of the $\left( \cdot ,\cdot \right) $ instruction scheme, has an infinite number of instructions which make it possible to observe the tree and modify it at arbitrary depths. However each individual instruction affects trees at a bounded depth, which is considered a constant time operation in functional language, which is the important property we want to ensure. As we shall see in Section~\ref{s:finitetype}, this choice of an infinite set of instruction is pure convenience: a finite set suffices.
  \par
  \subsection{A language of guards and actions}
  \par
  One way to read the instruction in $\Phi $, is to think of them as combinators giving means to match tree prefixes and rearrange the corresponding subtrees. That is, instructions of tree machines perform pattern-matching. We shall give an alternative set of instructions for tree machines which is suggestive of the pattern-matching notation functional programmers all know and love.
  \par
  We write $\gamma \Rightarrow \alpha $ for a partial function whose domain is denoted by the \emph{guard}, or pattern, $\gamma $ and whose functional action is denoted by the \emph{action} $\alpha $. Both $\gamma $ and $\alpha $ are trees with variables, with the following restrictions: variables occur at most once $\gamma $, and all variables of $\alpha $ appear in $\gamma $ (variables in $\gamma $ bind variables in $\alpha $). We may use ${\gamma }_{1}\Rightarrow {\alpha }_{1}~|~{\gamma }_{2}\Rightarrow {\alpha }_{2}$ instead of $\left( {\gamma }_{1}\Rightarrow {\alpha }_{1}\right) +\left( {\gamma }_{2}\Rightarrow {\alpha }_{2}\right) $ when ${\gamma }_{1}$ and ${\gamma }_{2}$ denote disjoint domains (\emph{i.e.} ${\gamma }_{1}$ and ${\gamma }_{2}$ are not unifiable). We can also use the wildcard pattern ``$\_$'' to represent a variable in $\gamma $ which does not bind a variable in $\alpha $.
  \par
  For instance, the instructions of $\Phi $ can be represented using this notations as follows:
  \begin{displaymath}
    \parbox{0.85\linewidth}{$
    \begin{array}{l@{\quad }c@{\quad }l@{\qquad }l}
      \mathbf{1} & = & x\Rightarrow x & \\
      \delta  & = & x\Rightarrow \left( x,x\right)  & \\
      {\pi }_{1} & = & \left( x,\_\right) \Rightarrow x & \\
      {\pi }_{2} & = & \left( \_,y\right) \Rightarrow y & \\
      \left( i,j\right)  & = & \left( {\gamma }_{i},{\gamma }_{j}\right) \Rightarrow \left( {\alpha }_{i},{\alpha }_{j}\right)  & (\mbox{for $i={\gamma }_{i}\Rightarrow {\gamma }_{j}$ and $j={\gamma }_{j}\Rightarrow {\alpha }_{j}$})\\
      \varepsilon  & = & \_\Rightarrow \left( \right)  & \\
      \left( \right)  & = & \left( \right) \Rightarrow \left( \right)  & \\
    \end{array}
    $}
  \end{displaymath}
  \par
  Conversely, the language of guard and action is subsumed by $\Phi $, which we shall use as a definition rather than giving and independent definition and prove it as a theorem (which would be, of course, theoretically possible but not practically useful). The definition is lexicographically recursive on the subterm ordering of $\gamma $ then that of $\alpha $.
  \begin{displaymath}
    \parbox{0.85\linewidth}{$
    \begin{array}{l@{\quad }c@{\quad }l@{\qquad }l}
      \left( x\Rightarrow x\right)  & = & \mathbf{1} & \\
      \left( x\Rightarrow \left( \right) \right)  & = & \varepsilon  & \\
      \left( \left( \right) \Rightarrow \left( \right) \right)  & = & \left( \right)  & \\
      \left( \left( {\gamma }_{1},{\gamma }_{2}\right) \Rightarrow \left( \right) \right)  & = & \left( \left( {\gamma }_{1}\Rightarrow \left( \right) \right) ,\left( {\gamma }_{2}\Rightarrow \left( \right) \right) \right) \cdot \varepsilon  & \\
      \left( \left( {\gamma }_{1},{\gamma }_{2}\right) \Rightarrow x\right)  & = & \left( \left( {\gamma }_{1}\Rightarrow x\right) ,\left( {\gamma }_{2}\Rightarrow \left( \right) \right) \right) \cdot {\pi }_{1} & (\mbox{when $x$ occurs in ${\gamma }_{1}$})\\
      \left( \left( {\gamma }_{1},{\gamma }_{2}\right) \Rightarrow x\right)  & = & \left( \left( {\gamma }_{1}\Rightarrow \left( \right) \right) ,\left( {\gamma }_{2}\Rightarrow x\right) \right) \cdot {\pi }_{2} & (\mbox{when $x$ occurs in ${\gamma }_{2}$})\\
      \left( \gamma \Rightarrow \left( {\alpha }_{1},{\alpha }_{2}\right) \right)  & = & \delta \cdot \left( \left( \gamma \Rightarrow {\alpha }_{1}\right) ,\left( \gamma \Rightarrow {\alpha }_{2}\right) \right)  & \\
    \end{array}
    $}
  \end{displaymath}
  The language of pattern and action can be used to conveniently defined the following examples:
  \begin{displaymath}
    \parbox{0.85\linewidth}{$
    \begin{array}{l@{\quad }c@{\quad }l@{\qquad }l}
      \sigma  & = & \left( \left( x,y\right) \Rightarrow \left( y,x\right) \right)  & \\
      \mbox{} & = & \delta \cdot \left( \left( \varepsilon ,\mathbf{1}\right) \cdot {\pi }_{2},\left( \mathbf{1},\varepsilon \right) \cdot {\pi }_{1}\right)  & \\
      \mathsf{push} & = & \left( \left( x,y\right) ,z\right) \Rightarrow \left( x,\left( y,z\right) \right)  & \\
      \mbox{} & = & \delta \cdot \left( \left( \left( \mathbf{1},\varepsilon \right) \cdot {\pi }_{1},\varepsilon \right) \cdot {\pi }_{1},\delta \cdot \left( \left( \left( \varepsilon ,\mathbf{1}\right) \cdot {\pi }_{2},\varepsilon \right) \cdot {\pi }_{1},\left( \left( \varepsilon ,\varepsilon \right) \cdot \varepsilon ,\mathbf{1}\right) \cdot {\pi }_{2}\right) \right)  & \\
    \end{array}
    $}
  \end{displaymath}
  This procedure does not give the smallest possible definition in terms of $\Phi $ of $\sigma $ and $\mathsf{push}$ (or of pretty much anything for that matter). Here are better candidates for this particular award:
  \begin{displaymath}
    \parbox{0.85\linewidth}{$
    \begin{array}{l@{\quad }c@{\quad }l@{\qquad }l}
      \sigma  & = & \delta \cdot \left( {\pi }_{2},{\pi }_{1}\right)  & \\
      \mathsf{push} & = & \delta \cdot \left( {\pi }_{1}\cdot {\pi }_{1},\delta \cdot \left( {\pi }_{2}\cdot {\pi }_{1},{\pi }_{2}\right) \right)  & \\
    \end{array}
    $}
  \end{displaymath}
  In either case, however, it is fair to claim that the language of guard and patterns gives a clearer account of the intent and semantics of instructions than the more elementary $\Phi $. In consequence we will peruse the guard and actions in the remainder of the article.
  \par
  \subsection{Constants}
  \par
  It will be useful to embed natural numbers in trees. Any embedding will do. We choose a binary encoding for the sake of fun and compactness:
  \begin{displaymath}
    \parbox{0.85\linewidth}{$
    \begin{array}{l@{\quad }c@{\quad }l@{\qquad }l}
      0 & = & \left( \right)  & \\
      2\times n+1 & = & \left( \left( \right) ,n\right)  & \\
      2\times n+2 & = & \left( \left( \left( \right) ,\left( \right) \right) ,n\right)  & \\
    \end{array}
    $}
  \end{displaymath}
  \par
  A direct consequence of this encoding is that any finite set of symbols can be easily represented as particular trees, by mapping them to arbitrary distinct natural numbers. We will do so quite liberally.
  \par
  \subsection{Zipper}\label{s:zipper}
  \par
  In~\cite{Huet1997} Huet presents a purely functional data structure, the \emph{zipper} to implement ``pointers'' in trees, \emph{i.e.} a way to walk through a tree from parent to child or back in constant time and to replace the pointed subtree also in constant time.
  \par
  The zipper can be adapted to the tree machines as a set of instructions. The zipper instruction provide us with an ability we have not had so far: modifying a subtree at an \emph{a priori} unbounded depth in a tree.
  \par
  The idea is that a zipper is represented as a pair $\left( c,t\right) $ of the pointed subtree $t$ together with a ``reversed tree'' $c$ which represents the context and gives enough information to rebuild the tree when walking up towards the parent (with the instruction $\mathsf{up}$ below). Here are the relevant instructions:
  \begin{displaymath}
    \parbox{0.85\linewidth}{$
    \begin{array}{l@{\quad }c@{\quad }l@{\qquad }l}
      \mathsf{open} & = & x\Rightarrow \left( \left( \right) ,x\right)  & \\
      \mathsf{left} & = & \left( x,\left( y,z\right) \right) \Rightarrow \left( \left( 0,\left( x,z\right) \right) ,y\right)  & \\
      \mathsf{right} & = & \left( x,\left( y,z\right) \right) \Rightarrow \left( \left( 1,\left( x,y\right) \right) ,z\right)  & \\
      \mathsf{up} & = & \left( \left( 0,\left( x,z\right) \right) ,y\right) \Rightarrow \left( x,\left( y,z\right) \right) ~|~\left( \left( 1,\left( x,y\right) \right) ,z\right) \Rightarrow \left( x,\left( y,z\right) \right)  & \\
      \mathsf{exit} & = & \left( \left( \right) ,x\right) \Rightarrow x & \\
    \end{array}
    $}
  \end{displaymath}
  Where $\mathsf{open}$ and $\mathsf{exit}$ transform a tree into a pointer to its root and back, $\mathsf{left}$ walks down to the left child (and marks, in the reversed tree, with $0$ that it did walk down left), $\mathsf{right}$ to the right child, and $\mathsf{up}$ walks up to the parent, and reconstruct the tree according to the mark left by either $\mathsf{left}$ or $\mathsf{right}$. For further detail, the reader unfamiliar with these concepts is deeply encouraged to read Huet's paper~\cite{Huet1997}
  \par
  \section{Comparisons}
  \par
  With the basic material now set in place, we can now turn to the use of the tree machine as a computational complexity model. It is important to be precise on what is meant here: clearly, complexity classes are very robust, and it does not matter what computation model is taken to define them; the tree machine is no exception. However, for more fine grained accounts of asymptotic complexity, the model will matter a lot: in Turing machines already, multiple-tape Turing machines can provide a quadratic speedup over single-tape ones.
  \par
  It is this sort of complexity that is our concern, and the claim of this article is that the tree machine is a good complexity model for purely functional computations.
  \par
  \subsection{Turing machines as tree machines}
  \par
  It is straightforward to implement Turing machines as tree-machines: fixing a coding for the alphabet, we arrange the tree to be a pair $\left( L,R\right) $ of lists of symbols. The list $R=\left( {\mathsf{a}}_{0},\left( {\mathsf{a}}_{1},\left( \mbox{\ldots },\left( \right) \mbox{\ldots }\right) \right) \right) $ represents the part of the tape just under and to the right of the head (${\mathsf{a}}_{0}$ is the symbol under the head). The reversed list $L=\left( \left( \left( \mbox{\ldots }\left( \right) ,\mbox{\ldots }\right) ,{\mathsf{b}}_{2}\right) ,{\mathsf{b}}_{1}\right) $ represents the part of the tape which sits to the left of the head. With this representation there is no need for a special symbol to stand at empty slots on the tape: instead the symbol under the head is empty when $R$ is the empty list $\left( \right) $.
  \par
  The instruction of Turing machines are implemented as instruction of the tree machines (in guard-and-action style):
  \begin{itemize}
    \item Write symbol $\mathsf{a}$ under the head: $\left( L,\left( \_,R\right) \right) \Rightarrow \left( L,\left( \mathsf{a},R\right) \right) ~|~\left( x,\left( \right) \right) \Rightarrow \left( x,\left( \mathsf{a},\left( \right) \right) \right) $ (the second case extends the tape if we reached the end)
    \item Move right: $\left( L,\left( x,R\right) \right) \Rightarrow \left( \left( L,x\right) ,R\right) $
    \item Move left: $\left( \left( L,x\right) ,R\right) \Rightarrow \left( L,\left( x,R\right) \right) $
    \item Check that symbol $\mathsf{a}$ is under the head: $\left( L,\left( \mathsf{a},R\right) \right) \Rightarrow \left( L,\left( \mathsf{a},R\right) \right) $
  \end{itemize}
  Therefore, a Turing machines is translated to a tree machine with the same state and transitions, except the instructions labelling transitions are replaced with their respective implementation as tree machine instructions.
  \par
  This translation highlights a point which is occasionally overlooked: Turing machines are not a very good model of imperative programs. Turing machines can be simulated in constant time in a purely functional language, and so can multiple-tape Turing machines. To get a better model of imperative programs, we shall turn to random-access machines in Section~\ref{s:ram}.
  \par
  \subsection{A finite type for tree machines}\label{s:finitetype}
  \par
  In order to build the converse translation of tree machines into Turing machines, it will be convenient for the set instruction to be presented by a finite set $\Psi $. To obtain this finite presentation we will use the zipper instructions from Section~\ref{s:zipper}. To be more specific, we will implement the instruction $\left( i,j\right) \in \Phi $ as a (finite) sequence of zipper operations which will walk through the tree to apply the appropriate actions to the appropriate subtrees.
  \par
  Remember that a zipper is a pair $\left( c,t\right) $ of a focused subtree $t$ together with a reversed tree $c$ representing the necessary context to rebuild the tree. Our goal is to lift the instructions of $\Phi $ so that they apply to the focused subtree instead of the root of the complete tree. In other words, we are looking for a $\llbracket i\rrbracket =\left( 1,i\right) $ for each $i\in \Phi $.
  \par
  This property that $\llbracket i\rrbracket =\left( 1,i\right) $ acts as a perfectly fine definition for each of the generators of $\Phi $:
  \begin{displaymath}
    \parbox{0.85\linewidth}{$
    \begin{array}{l@{\quad }c@{\quad }l@{\qquad }l}
      \llbracket \mathbf{1}\rrbracket  & = & \left( \mathbf{1},\mathbf{1}\right)  & \\
      \llbracket \left( \right) \rrbracket  & = & \left( \mathbf{1},\left( \right) \right)  & \\
      \llbracket \varepsilon \rrbracket  & = & \left( \mathbf{1},\varepsilon \right)  & \\
      \llbracket {\pi }_{1}\rrbracket  & = & \left( \mathbf{1},{\pi }_{1}\right)  & \\
      \llbracket {\pi }_{2}\rrbracket  & = & \left( \mathbf{1},{\pi }_{2}\right)  & \\
      \llbracket \delta \rrbracket  & = & \left( \mathbf{1},\delta \right)  & \\
    \end{array}
    $}
  \end{displaymath}
  For the case $\left( i,j\right) $ however, in order to avoid introducing infinitely many instructions in $\Psi $, we need to find an alternative definition in terms of the zipper operations. We define $\llbracket \left( i,j\right) \rrbracket $ recursively as follows:
  \begin{displaymath}
    \parbox{0.85\linewidth}{$
    \begin{array}{l@{\quad }c@{\quad }l@{\qquad }l}
      \llbracket \left( i,j\right) \rrbracket  & = & \mathsf{left}\cdot \llbracket i\rrbracket \cdot \mathsf{up}\cdot \mathsf{right}\cdot \llbracket j\rrbracket \cdot \mathsf{up} & \\
    \end{array}
    $}
  \end{displaymath}
  It is a straightforward exercise of symbol pushing to verify that indeed, by induction, $\llbracket \left( i,j\right) \rrbracket =\left( \mathbf{1},\left( i,j\right) \right) $.
  \par
  Every instruction $i\in \Phi $ can be implemented as $\mathsf{open}\cdot \llbracket i\rrbracket \cdot \mathsf{exit}$. Therefore we can take the set $\Phi $ as being
  \begin{displaymath}
    \Psi =\left\{ \mathsf{open};\mathsf{left};\mathsf{right};\mathsf{up};\mathsf{exit};\left( \mathbf{1},\mathbf{1}\right) ;\left( \mathbf{1},\left( \right) \right) ;\left( \mathbf{1},\varepsilon \right) ;\left( \mathbf{1},{\pi }_{1}\right) ;\left( \mathbf{1},{\pi }_{2}\right) ;\left( \mathbf{1},\delta \right) \right\} 
  \end{displaymath}
  \par
  \subsection{Tree machines as Turing machines}\label{s:to:turing}
  \par
  Translating tree machines into Turing machines is not as direct as the converse. One way to translate trees into word so that it fits a Turing machine tape is to use the Polish notations: we take the alphabet to include $\left\{ \mathsf{p};\mathsf{u}\right\} $ (for \emph{pair} and \emph{unit} respectively). The tree $\left( \left( \right) ,\left( \left( \right) ,\left( \right) \right) \right) $ is then translated to $\mathsf{pupuu}$.
  \par
  As always, giving a concrete definition of a Turing machine~--~or even of a translation to Turing machine~--~is not very easy nor particularly enlightening, and, in fact, would lead us way over the page limit. On the other hand, equipped with the finite presentation of Section~\ref{s:finitetype}, it is quite clear that it can in principle be done.
  \par
  Let us sketch how such a construction could be achieved. In a first step we can ignore the $\mathsf{open}$ and $\mathsf{exit}$ instruction, and assume we are always working on a zipper. The rational is that $\mathsf{exit}\cdot \mathsf{open}$ acts as the identity on a zipper which is focused at the root, so we may simply represent non-zipper trees as zipper focused at the root and both instructions become the identity. One may be tempted to replace calls to $\mathsf{exit}$ to test that the context is a $\mathsf{u}$, but this is not even necessary as it is an invariant of the translation.
  \par
  To represent the zipper $\left( c,t\right) $, the most convenient way is to have two tapes, one holding $c$ and the other holding $t$. A third tape will be used to store a counter to navigate through Polish-notation trees, and a fourth to store intermediate trees which are to be moved or copied.
  \par
  The instructions of tree machines are not translated as constant time instructions. However, they are all in a polynomial $P$ (which depends on the details of the translation, but is at least of degree $1$) of the current size of the tape. Hence, if the complexity of a tree machine is $O{\left( f{\left( n\right) }\right) }$, then the corresponding Turing machine has complexity $O{\left( P{\left( f{\left( n\right) }\right) }\times f{\left( n\right) }\right) }$, which is in the same complexity class.
  \par
  Therefore, tree machines and Turing machines have the same complexity classes. However, the translation from tree machines to Turing machines is non-trivial both in term of slowdowns of the translated machine and complexity of the translation itself. It would be quite hard to get an explicit description of the translation. On the other hand the translation of Turing machines into tree machines is quite direct. Tree machines fare pretty well on that front.
  \par
  \subsection{Tree machines and random-access machines}\label{s:ram}
  \par
  Random-access machines~--~which happen to be yet another instance of Eilenberg machine~--~support more natural translations of tree machines: the encoding of algebraic data types of functional programming languages. In such a translation, a tree is encoded as an address, at this address there is $0$ if the tree is empty, and $1$ if the tree is a pair. In the latter case, the two following addresses contain the addresses of the two subtrees.
  \par
  Unfortunately, this translation would not fit these pages either, as it has to solve the problem of memory allocation and garbage collection in order to preserve the space complexity of tree machines. Garbage collection can be done achieved via reference counting~\cite{Wilson1992} since the tree cannot have cycles, but it still would not make a program under a page long (or anything near it).
  \par
  Nevertheless, this translation is quite concrete and serves as a good test for the tree machine. Tree machines, indeed, can implemented in the traditional complexity model of practical computer and, assuming garbage collection away, this implementation preserves the complexity of machines. Assuming that garbage collection is constant time may seem unreasonable, but it is in fact the way functional programmers think about their programs: as having negligible overhead due to garbage collection.
  \par
  This translation also serves to remark a limitation of the tree machine for space efficiency. Indeed trees can have various representations in a random-access machine with more or less sharing between subtrees. In the worst case, a maximally shared subtree is exponentially smaller than its sharing-free equivalent. Therefore, in an accurate cost model for tree-machine space consumption, the space occupied by a tree cannot be read on the tree itself: it depends on the history of how the tree was built. There is no particular problem in defining such a dynamic space-cost semantics though, see~\cite{Blelloch2013} for a much more ambitious case. It also means that the translation to Turing machines in Section~\ref{s:to:turing} is not accurate as far as space complexity is concerned.
  \par
  Conversely, it is not difficult to implement a random-access machine as a tree machine. Random-access machines are composed of arithmetic operations and an addressed memory. Arithmetic is straightforward, and addressed memory can be stored in a tree (most likely a trie such as in~\cite[Chapter 10]{Okasaki1999}).
  \par
  Random-access machines typically assume constant time arithmetic operations and memory access. The translation of random-access machines into tree machines preserves neither. For the case of constant-time arithmetic, the assumption is not realistic for big numbers, it is actually meant to model the fact that no big numbers will appear and that bounded arithmetic is sufficient for the modelled algorithms. To make such an hypothesis in tree machines, they would need to be outfitted with a primitive notion of integers like random-access machines.
  \par
  Constant-time memory access is more subtle: it is a reasonable assumption for a vast majority of programs, yet, a more accurate model may use logarithmic-time memory access~\cite{Aggarwal1987}. Logarithmic-time memory access is precisely what is provided by the tree machine, so it might look as though the superiority of the random-access machine were illusory. It is not, however, as when a model with logarithmic-time memory access apply, it will not be \emph{a priori} reasonable to see the instructions of the tree machines as running in constant time.
  \par
  \section{Encoding ${\lambda}$-calculus}\label{s:lambda}
  \par
  To conclude, I want to present, with few words, an encoding of ${\lambda}$-calculus in tree machines to demonstrate that it is really practical to make explicit definitions of non-trivial machines.
  \par
  We shall use the simple explicit substitution calculus called ${\lambda}$${\upsilon}$-calculus~\cite{Lescanne1994}. It uses de Bruijn indices and has three kinds of substitutions: $[v]$ for a term $v$, $[\uparrow ]$ (read \emph{shift}) and $[\Uparrow s]$ (read \emph{lift $s$}) for a substitution $s$.
  \begin{displaymath}
    \parbox{0.85\linewidth}{$
    \begin{array}{l@{\quad }c@{\quad }l@{\qquad }l}
      (\lambda u)v & \leadsto  & u[v] & \\
      (\lambda u)[s] & \leadsto  & \lambda (u[\Uparrow s]) & \\
      (u v)[s] & \leadsto  & (u[s])(v[s]) & \\
      0[v] & \leadsto  & v & \\
      (n+1)[v] & \leadsto  & n+1 & \\
      0[\Uparrow s] & \leadsto  & 0 & \\
      (n+1)[\Uparrow s] & \leadsto  & n[s][\uparrow ] & \\
      n[\uparrow ] & \leadsto  & n+1 & \\
    \end{array}
    $}
  \end{displaymath}
  \par
  Let us suppose fixed an encoding for the symbols that we will use in the encoding of terms: $\left\{ \lambda ;\mathsf{app};\mathsf{succ};\mathsf{nought};\mathsf{subst};\mathsf{term};\mathsf{lift};\mathsf{shift};\mathsf{ok}\right\} $. The terms $\lambda u$ and $u v$ are represented, respectively $\left( \lambda ,u\right) $ and $\left( \mathsf{app},\left( u,v\right) \right) $. The de Bruijn index $2$ is represented as $\left( \mathsf{succ},\left( \mathsf{succ},\mathsf{nought}\right) \right) $.
  \par
  We will use the non-determinism of tree machines to represent the non-determinism of ${\beta}$-reduction, which is achieved navigating non-deterministically in ${\lambda}$-terms with zipper-like instructions.
  \begin{displaymath}
    \parbox{0.85\linewidth}{$
    \begin{array}{l@{\quad }c@{\quad }l@{\qquad }l}
      \mathsf{open} & = & u\Rightarrow \left( \left( \right) ,u\right)  & \\
      {\mathsf{down}}_{\lambda } & = & \left( c,\left( \lambda ,u\right) \right) \Rightarrow \left( \left( c,\lambda \right) ,u\right)  & \\
      {\mathsf{down}}_{\sigma } & = & \left( c,\left( \mathsf{subst},\left( u,s\right) \right) \right) \Rightarrow \left( \left( c,\left( \mathsf{subst},s\right) \right) ,u\right)  & \\
      \mathsf{left} & = & \left( c,\left( \mathsf{app},\left( u,v\right) \right) \right) \Rightarrow \left( \left( c,\left( \mathsf{app},\left( 0,v\right) \right) \right) ,u\right)  & \\
      \mathsf{right} & = & \left( c,\left( \mathsf{app},\left( u,v\right) \right) \right) \Rightarrow \left( \left( c,\left( \mathsf{app},\left( 1,u\right) \right) \right) ,v\right)  & \\
    \end{array}
    $}
  \end{displaymath}
  \begin{displaymath}
    \parbox{0.85\linewidth}{$
    \begin{array}{l@{\quad }c@{\quad }l@{\qquad }l}
      {\mathsf{up}}_{\lambda } & = & \left( \left( c,\lambda \right) ,u\right) \Rightarrow \left( c,\left( \lambda ,u\right) \right)  & \\
      {\mathsf{up}}_{\sigma } & = & \left( \left( c,\left( \mathsf{subst},s\right) \right) ,u\right) \Rightarrow \left( c,\left( \mathsf{subst},\left( u,s\right) \right) \right)  & \\
      {\mathsf{up}}_{{\mathsf{app}}_{0}} & = & \left( \left( c,\left( \mathsf{app},\left( 0,v\right) \right) \right) ,u\right) \Rightarrow \left( c,\left( \mathsf{app},\left( u,v\right) \right) \right)  & \\
      {\mathsf{up}}_{{\mathsf{app}}_{1}} & = & \left( \left( c,\left( \mathsf{app},\left( 1,u\right) \right) \right) ,v\right) \Rightarrow \left( c,\left( \mathsf{app},\left( u,v\right) \right) \right)  & \\
      \mathsf{up} & = & {\mathsf{up}}_{\lambda }~|~{\mathsf{up}}_{\sigma }~|~{\mathsf{up}}_{{\mathsf{app}}_{0}}~|~{\mathsf{up}}_{{\mathsf{app}}_{1}} & \\
      \mathsf{exit} & = & \left( \left( \right) ,u\right) \Rightarrow u & \\
      \mathsf{move} & = & {\mathsf{down}}_{\lambda }+{\mathsf{down}}_{\sigma }+\mathsf{left}+\mathsf{right}+\mathsf{up} & \\
      \mathsf{zip} & = & {\mathsf{up}}^{*}\cdot \mathsf{exit} & \\
    \end{array}
    $}
  \end{displaymath}
  The reduction rules are given in a context independent manner:
  {\small{
  \begin{displaymath}
    \parbox{0.85\linewidth}{$
    \begin{array}{l@{\quad }c@{\quad }l@{\qquad }l}
      \beta  & = & \left( \mathsf{app},\left( \left( \lambda ,u\right) ,v\right) \right) \Rightarrow \left( \mathsf{subst},\left( u,\left( \mathsf{term},v\right) \right) \right)  & \\
      {\lambda }_{\sigma } & = & \left( \mathsf{subst},\left( \left( \lambda ,u\right) ,s\right) \right) \Rightarrow \left( \lambda ,\left( \mathsf{subst},\left( u,\left( \mathsf{lift},s\right) \right) \right) \right)  & \\
      {\mathsf{app}}_{\sigma } & = & \left( \mathsf{subst},\left( \left( \mathsf{app},\left( u,v\right) \right) ,s\right) \right) \Rightarrow \left( \mathsf{app},\left( \left( \mathsf{subst},\left( u,s\right) \right) ,\left( \mathsf{subst},\left( v,s\right) \right) \right) \right)  & \\
      {\mathsf{nought}}_{\mathsf{term}} & = & \left( \mathsf{subst},\left( \mathsf{nought},\left( \mathsf{term},v\right) \right) \right) \Rightarrow v & \\
      {\mathsf{succ}}_{\mathsf{term}} & = & \left( \mathsf{subst},\left( \left( \mathsf{succ},n\right) ,\left( \mathsf{term},\_\right) \right) \right) \Rightarrow \mathsf{succ}{\left( n\right) } & \\
      {\mathsf{nought}}_{\mathsf{lift}} & = & \left( \mathsf{subst},\left( \mathsf{nought},\left( \mathsf{lift},\_\right) \right) \right) \Rightarrow \mathsf{nought} & \\
      {\mathsf{succ}}_{\mathsf{lift}} & = & \left( \mathsf{subst},\left( \left( \mathsf{succ},n\right) ,\left( \mathsf{lift},s\right) \right) \right) \Rightarrow \left( \mathsf{subst},\left( \left( \mathsf{subst},\left( x,s\right) \right) ,\mathsf{shift}\right) \right)  & \\
      {\mathsf{nought}}_{\mathsf{shift}} & = & \left( \mathsf{subst},\left( \mathsf{nought},\mathsf{shift}\right) \right) \Rightarrow \left( \mathsf{succ},\mathsf{nought}\right)  & \\
      {\mathsf{succ}}_{\mathsf{shift}} & = & \left( \mathsf{subst},\left( \left( \mathsf{succ},n\right) ,\mathsf{shift}\right) \right) \Rightarrow \left( \mathsf{succ},\left( \mathsf{succ},n\right) \right)  & \\
      {\mathsf{var}}_{\mathsf{shift}} & = & {\mathsf{nouhgt}}_{\mathsf{shift}}~|~{\mathsf{succ}}_{\mathsf{shift}} & \\
      \sigma  & = & {\lambda }_{\sigma }~|~{\mathsf{app}}_{\sigma }~|~{\mathsf{nought}}_{\mathsf{term}}~|~{\mathsf{succ}}_{\mathsf{term}}~|~{\mathsf{nought}}_{\mathsf{lift}}~|~{\mathsf{succ}}_{\mathsf{lift}}~|~{\mathsf{var}}_{\mathsf{shift}} & \\
    \end{array}
    $}
  \end{displaymath}
  }}
  A step of ${\beta}$-reduction is represented as one ${\beta}$ rule followed by eagerly applying ${\sigma}$ rules. To ensure that explicit substitutions have been eliminated, substitution-free terms are marked with $\mathsf{ok}$.
  \begin{displaymath}
    \parbox{0.85\linewidth}{$
    \begin{array}{l@{\quad }c@{\quad }l@{\qquad }l}
      {\mathsf{nought}}_{\mathsf{ok}} & = & \mathsf{nought}\Rightarrow \left( \mathsf{ok},\mathsf{nought}\right)  & \\
      {\mathsf{succ}}_{\mathsf{ok}} & = & \left( \mathsf{succ},n\right) \Rightarrow \left( \mathsf{ok},\left( \mathsf{succ},n\right) \right)  & \\
      {\lambda }_{\mathsf{ok}} & = & \left( \lambda ,\left( \mathsf{ok},u\right) \right) \Rightarrow \left( \mathsf{ok},\left( \lambda ,u\right) \right)  & \\
      {\mathsf{app}}_{\mathsf{ok}} & = & \left( \mathsf{app},\left( \left( \mathsf{ok},u\right) ,\left( \mathsf{ok},v\right) \right) \right) \Rightarrow \left( \mathsf{ok},\left( \mathsf{app},\left( u,v\right) \right) \right)  & \\
      {\mathsf{rule}}_{\mathsf{ok}} & = & {\mathsf{nought}}_{\mathsf{ok}}+{\mathsf{succ}}_{\mathsf{ok}}+{\lambda }_{\mathsf{ok}}+{\mathsf{app}}_{\mathsf{ok}} & \\
      {\mathsf{check}}_{\mathsf{ok}} & = & \left( \mathsf{ok},u\right) \Rightarrow u & \\
    \end{array}
    $}
  \end{displaymath}
  A step of ${\beta}$-reduction is encoded as the following machine\footnote{Notations are abused a little in this example, as $\left( \cdot ,\cdot \right) $ is used with relations which are not instructions from the type $\Phi $. It can be made formal using Section~\ref{s:finitetype}.}:
  \begin{displaymath}
    \parbox{0.85\linewidth}{$
    \begin{array}{l@{\quad }c@{\quad }l@{\qquad }l}
      {\mathsf{all}}_{\sigma } & = & {\left( {\mathsf{move}}^{*}\cdot \left( \mathbf{1},\sigma \right) \right) }^{*} & \\
      {\mathsf{zip}}_{\mathsf{ok}} & = & {\left( {\mathsf{move}}^{*}\cdot \left( \mathbf{1},{\mathsf{rule}}_{\mathsf{ok}}\right) \right) }^{*}\cdot \mathsf{exit}\cdot {\mathsf{check}}_{\mathsf{ok}} & \\
      \mathsf{step} & = & \mathsf{open}\cdot {\mathsf{move}}^{*}\cdot \left( \mathbf{1},\beta \cdot \mathsf{open}\cdot {\mathsf{all}}_{\sigma }\cdot {\mathsf{zip}}_{\mathsf{ok}}\right) \cdot \mathsf{zip} & \\
    \end{array}
    $}
  \end{displaymath}
  \section{Conclusion}
  \par
  The tree machine is designed to stand as a standard model of complexity for purely functional algorithms, much like the random-access machine is for imperative algorithms. At the cost of not being quite as minimalist as Turing machines, tree machines are quite expressive and make it reasonably easy to write explicit machines to implement desired behaviours.
  \par
  In fact, if in this article I have given direct descriptions of machines, a technique, favoured by Danvy, allows to compile programs written in a very small purely functional idealised Scheme language, with full recursion and higher order, into a tree machine. This is achieved by apply sequentially to the program the following transformations: defunctionalisation, continuation-passing style transformation, and defunctionalisation again~\cite{Ager2003}. This sequence of transformation yields a mutually recursive first-order program where all calls (in particular recursive calls) are tail, while preserving the structure of the program. At this point we are one tiny step shy of having a tree machine: function environments must be reified as data and variables as accessors in this data. Then, using the guard-and-action presentation of tree machines, we can build a machine with just one state per function in the transformed program (the functions of the original program plus one or two administrative functions). Yielding another, more indirect, way to produce explicit tree machines.
  \bibliography{library}

\begin{thebibliography}{1}

\bibitem{Ager2003}
Mads~Sig Ager, Dariusz Biernacki, Olivier Danvy, and Jan Midtgaard.
\newblock {A functional correspondence between evaluators and abstract
  machines}.
\newblock In {\em Proceedings of the 5th ACM SIGPLAN international conference
  on Principles and practice of declaritive programming - PPDP '03}, number
  March, pages 8--19, New York, New York, USA, 2003. ACM Press.

\bibitem{Aggarwal1987}
Alok Aggarwal, Bowen Alpern, Ashok~K Chandra, and Marc Snir.
\newblock {A model for hierarchical memory}.
\newblock {\em Proceedings of the nineteenth annual ACM symposium on Theory of
  computing}, pages 305----314, 1987.

\bibitem{Blelloch2013}
Guy Blelloch and Robert Harber.
\newblock {Cache and I/O efficent functional algorithms}.
\newblock In {\em Proceedings of the 40th annual ACM SIGPLAN-SIGACT symposium
  on Principles of programming languages - POPL '13}, page~39, New York, New
  York, USA, 2013. ACM Press.

\bibitem{Eilenberg1974}
Samuel Eilenberg.
\newblock {\em {Automata, languages, and machines}}.
\newblock 1974.

\bibitem{Huet1997}
G\'{e}rard Huet.
\newblock {Functional Pearl: The Zipper}.
\newblock {\em Journal of Functional Programming}, 7:7--5, 1997.

\bibitem{Lescanne1994}
Pierre Lescanne.
\newblock {From $\lambda$$\sigma$ to $\lambda$$\upsilon$: a journey through
  calculi of explicit substitutions}.
\newblock In {\em Proceedings of the 21st ACM SIGPLAN-SIGACT symposium on
  Principles of programming languages - POPL '94}, pages 60--69, New York, New
  York, USA, 1994. ACM Press.

\bibitem{Okasaki1999}
Chris Okasaki.
\newblock {\em {Purely functional data structures}}.
\newblock Cambridge University Press, 1999.

\bibitem{Wilson1992}
Paul~R Wilson.
\newblock {Uniprocessor garbage collection techniques}.
\newblock {\em Memory Management}, 1992.

\end{thebibliography}
\end{document}